\documentclass[11pt]{article}
\usepackage{setspace}
\usepackage{blindtext, graphicx}
\usepackage{graphicx}
\usepackage{multirow}
\usepackage{floatrow}
\usepackage[T1]{fontenc}
\usepackage{capt-of}
\usepackage{enumitem}
\DeclareFloatFont{tiny}{\scriptsize}%
\usepackage[utf8]{inputenc}
\usepackage{dtklogos} 
\usepackage{caption}
\usepackage{subfigure}
\usepackage{mathtools}
\usepackage{listings}
\usepackage{color}
\usepackage{soul}
\usepackage{array}
\newcolumntype{P}[1]{>{\centering\arraybackslash}p{#1}}
\usepackage{setspace} 
\usepackage{graphicx}
\usepackage{algcompatible}
\usepackage{afterpage,lipsum}
\usepackage{tikz}
\usepackage{longtable}
\usepackage{multirow}
\usepackage{pbox}
\usepackage{url}
\usepackage{lscape}
\usepackage{verbatim} 
\usepackage{array}
\usepackage{booktabs}
\usepackage{adjustbox,lipsum}
\usepackage[ruled,vlined,linesnumbered]{algorithm2e}
\usepackage{subfig} 
\usepackage{caption}
\usepackage{amsmath}
\usepackage{amsmath}
\usepackage{mathtools}
\usepackage{anyfontsize}

\usepackage{multicol}

\newcommand{\specialcell}[2][c]{%
  \begin{tabular}[#1]{@{}c@{}}#2\end{tabular}}

\usepackage{etoolbox,refcount}
\usepackage{multicol}

\newcounter{countitems}
\newcounter{nextitemizecount}
\newcommand{\setupcountitems}{%
  \stepcounter{nextitemizecount}%
  \setcounter{countitems}{0}%
  \preto\item{\stepcounter{countitems}}%
}
\makeatletter
\newcommand{\computecountitems}{%
  \edef\@currentlabel{\number\c@countitems}%
  \label{countitems@\number\numexpr\value{nextitemizecount}-1\relax}%
}
\newcommand{\nextitemizecount}{%
  \getrefnumber{countitems@\number\c@nextitemizecount}%
}
\newcommand{\previtemizecount}{%
  \getrefnumber{countitems@\number\numexpr\value{nextitemizecount}-1\relax}%
}
\makeatother    
{\end{itemize}%
\unskip\computecountitems\ifnumcomp{\previtemizecount}{>}{3}{\end{multicols}}{}}

\definecolor{dkgreen}{rgb}{0,0.6,0}
\definecolor{gray}{rgb}{0.5,0.5,0.5}
\definecolor{mauve}{rgb}{0.58,0,0.82}
\begin{document}
\title{Graph or Relational Databases: A Speed Comparison for Process Mining Algorithm} 
\author{Jeevan Joishi \quad Ashish Sureka\\
Indraprastha Institute of Information Technology, Delhi (IIITD)\\
       New Delhi, India\\
ABB Corporate Research\\
       Bangalore, India
}

\maketitle

%

\begin{abstract}
Process-Aware Information System (PAIS) are IT systems that manages, supports business processes and generate large event logs from execution of business processes. An event log is represented as a tuple of the form CaseID, TimeStamp, Activity and Actor. \emph{Process Mining} is an emerging area of research that deals with the study and analysis of business processes based on event logs. Process Mining aims at analyzing event logs and discover business process models, enhance them or check for conformance with an a priori model. The large volume of event logs generated are stored in databases. Relational databases perform well for certain class of applications. However, there are certain class of applications for which relational databases are not able to scale. A number of NoSQL databases have emerged to encounter the challenges of scalability. Discovering social network from event logs is one of the most challenging and important Process Mining task. Similar-Task and Sub-Contract algorithms are some of the most widely used Organizational Mining techniques. Our objective is to investigate which of the databases (Relational or Graph) perform better for Organizational Mining under Process Mining. An intersection of Process Mining and Graph Databases can be accomplished by modelling these Organizational Mining metrics with graph databases. We implement Similar-Task and Sub-Contract algorithms on relational and NoSQL (graph-oriented) databases using only query language constructs. We conduct empirical analysis on a large real world data set to compare the performance of row-oriented database and NoSQL graph-oriented  database. We benchmark performance factors like query execution time, CPU usage and  disk/memory space usage for NoSQL graph-oriented database against row-oriented database.
\end{abstract}
~

%

\section{Research Motivation and Aim}
\par{PAIS like ERP, CRM, etc. are IT systems that manages and supports business processes. The data generated by execution of activities within PAIS is in the form of event logs (tuple of the form <CaseID, TimeStamp, Activity, Actor>). An event log contains information on the business process being considered (CaseID), the set of events (Activities) within that CaseID, performer of the Activity (Actor) besides other information like TimeStamp and unique identifier. Process Mining is a area of research that aims on analyzing business processes based on event logs \cite{aalstmain}. Insights gathered from the analysis can be used by organizations to identify bottlenecks if any, improve or enhance their business process. For example, in the domain of Software Engineering, Gupta et al. mine bug report history for discovering process maps, inefficiencies and inconsistencies \cite{mgupta2014}. Based on whether an a priori model exists or not, Process Mining is classified into three broad techniques \textit{viz.} Process Discovery, Process Conformnce and Process Enhancement. Process Mining is divided into three different perspectives viz. Control Flow, Organizational and Case, based on the type of  attribute being considered from the event log \cite{aalstmain}. Control Flow perspective focuses on the lineage of business processes. Organizational Mining perspective deals with techniques used to study social structure within an organization \cite{aalst}, \cite{mxml}. Whereas Case perspective focuses on mining information within each process instance (CaseID). Organizational Mining is a Process Discovery technique which focuses on finding social networks between Actors of the event log. Various metrics for finding such sociograms are defined in \cite{aalst}.}
\par{Organizations have generally used Relational Databases (RDBMS) to store data. RDBMS handle tabular structures exceedingly well \cite{graphbook}. RDBMS generally focuses on Online Transaction Processing (OLTP) applications but are not found to be efficient for certain Online Analytical Processing (OLAP) applications that involve joins or analytical functions (Dense\textunderscore Rank, Sum, Average, etc.) at large scale. Developers have faced problems in trying to handle relationships with RDBMS mostly due to join intensive queries leading to JOIN BOMB\footnote{http://neo4j.com/blog/demining-the-join-bomb-with-graph-queries/}. The reason is that relationships in RDBMS can be modeled by means of joins only, and an increase in connectedness of data implies increased number of joins. Join intensive queries are an impediment to performance and scalability in a dynamic system with ever-changing business needs. Furthermore, complications arise when, in addition to modeling the relationships, we also need to weigh the strength of these relationships \cite{graphbook}.}
\par{Recent trends in database technologies has seen the emergence of various NoSQL databases. These databases breaks away from the traditional \textit{one-size-fits-all} philosophy employed by RDBMS and instead focuses on specific use cases \cite{one}. One such type of databases is Graph Database that are built to cater to linked data commonly found on social networking sites like Linkedin\footnote{https://www.linkedin.com/}, Facebook\footnote{https://www.facebook.com/}, etc. Graph databases have emerged to address the issue of leveraging complex and dynamic relationships in highly connected data. In contrast to relational databases, where performance deteriorates as the size of the dataset increases, performance of a graph database is expected to remain constant, even as the dataset grows \cite{graphbook}. This is because queries would be localized to a portion of the graph and hence, the execution time for each query would depend only on the part of the graph traversed to satisfy that query, instead of the overall size of the graph.}
\par{A lot of research has has been done on integrating data mining techniques directly into the DBMS \cite{ordonez2004}, \cite{ordonezc}, \cite{ordonezmcmc}. This allows for better data management, allows primitives to be defined at database levels and the applications are tightly coupled to the database. We aim to implement Organizational Mining algorithms \textit{viz.} Similar-Task and Sub-Contract, using only database language constructs and make these applications tightly coupled to the database. In view of the current work, our aim of this study can be summarized as-}
\setlist[1]{itemsep=-3pt}
\begin{enumerate}
\item To investigate approaches to transform Similar-Task and Sub-Contract algorithm in row-oriented database MySQL\footnote{http://www.mysql.com/}.
\item To examine approaches to adapt Similar-Task and Sub-Contract algorithm in graph-oriented database Neo4j\footnote{http://neo4j.com/}.
\item To conduct a series of experiments to benchmark and compare performance of Similar-Task and Sub-Contract algorithms in Neo4j against MySQL.
\end{enumerate}

%

\section{Related Work and Research Contribution}
In this Section, we closely review the related work to the study that are presented in this paper and also list the novel contributions of our work in context to existing work.
\normalfont \subsection {\normalfont \textbf{Implementation of Mining Algorithms in Relational Databases}}
\par{Ordonez et al. did an extensive work on implementing k-means clustering algorithm in SQL \cite{ordonez2004}. They came up three different SQL implementations of k-means algorithm to integrate it with RDBMS. Experiments were performed on large clusters, efficient indexing and with queries optimized and re-written.  Ordonez et al. also presented SQL implementations of EM Algorithm that worked with high dimensional data, high number of clusters and very large datasets \cite{ordonezc}. They came up with three different strategies viz. Horizontal, Vertical and Hybrid.  Ordonez et al. came up with another SQL implementation of clustering algorithm which merges Markov Chain Monte Carlo with EM algorithm \cite{ordonezmcmc}.  Sattler et al. described primitives for applying and building decision tree classifiers which were directly coupled on commercial databases used in various classification problems \cite{ron1996power}.}
\normalfont \subsection {\normalfont \textbf{Implementation of Mining Algorithms in Graph Databases}}
\par{Wang et al. presented papers that studied structural pattern mining for large disk based graph databases They presented a novel ADI index structure and efficient algorithms for mining frequent patterns \cite{wang1}. Wang et al again came up with novel techniques to obtain scalable mining on large disk based graph databases \cite{wang2}. Huan et al. also presented techniques to find maximal frequent sub-graphs from Graph Databases \cite{huan}. Ozaki came up with the concept of hyperclique pattern in graph databases to detect highly correlated sub-graph in graph structured databases. It considers general ordering of sub-graphs and employed techniques like breadth-first search/ depth-first search with powerful pruning techniques based on various measures \cite{ozaki}.}
\normalfont \subsection {\normalfont \textbf{Performance comparison between Relational and Graph Databases.}}
\par{Vicknair et al. performed comparisons between Relational Databases and Graph Databases. Their work included recording and querying data provenance information \cite{vicknair}. McColl et al. evaluated performance for a series of open-source graph databases. They used four different graph algorithms to evaluate performance for graph setup consisting upto 256 million nodes \cite{mccoll}. Ciglan et al. came up with benchmarking of graph databases over graph traversal algorithms \cite{ciglan}. Macko et al. presented a  performance introspection framework for graph databases, PIG. PIG provided techniques and tools to understand performance of graph databases \cite{macko}.}
\normalfont \subsection {\normalfont \textbf{Performance Analysis of Process Mining Algorithm on other Architecture}}
\par{Kundra et al. investigate the application of parallelization on Alpha Miner algorithm and use Graphics Processor Unit (GPU) to run computationally intensive parts of Alpha Miner algorithm in parallel. They demonstrate a highest speedup on GPU reaching 39-40 times from the same program run over multi-core CPU \cite{ kundra2016}. Sachdev et al. \cite{ sachdev2015}. Sachdev et al. investigate which of the databases (Relational or NoSQL) performs better for a Process Discovery application under Process Mining \cite{sachdev2015}. They conduct a performance benchmarking and comparison of the alpha-miner algorithm on row-oriented database and NoSQL column-oriented database \cite{sachdev2015}\cite{gupta2015}. Anand et al. Anand et al propose a Utility-Based Fuzzy Miner (UBFM) algorithm to efficiently mine a process model driven by a utility threshold and conduct experimental analysis to show the performance of the process mining algorithm on relational databases \cite{ anand2015}.}
\subsection{Novel Contributions}
\par{In context to existing work, the study presented in this paper makes the following novel contributions. The work presented in this paper is an \textbf{extension and detailed version of the paper by Joishi et al.} \cite{joishi2015}
\begin{enumerate}
\item While there has been work done on implementing data mining algorithms in row-oriented databases, we are the the first to implement Organizational Mining algorithms in relational databases.
\item While data mining algorithms like frequent pattern mining have been implemented in graph databases, we believe we are the first to implement Organizational Mining algorithms in graph databases.
\item We conduct a series of experiments to compare performance and benchmark Organizational Mining algorithms on graph databases against relational databases.
\end{enumerate}}

%

\section{Similar-Task and Sub-Contract Algorithm}
\par{An example  of an event log is shown in Table 1. Each row of the table is an event with CaseID, corresponding Activity and the Actor performing that Activity. We suggest readers to refer \cite{aalst}, \cite{mxml} for better understanding of Organizational Mining metrics.}
\renewcommand{\arraystretch}{.5}
\begin{table}
\CenterFloatBoxes
\begin{floatrow}
\ttabbox
  { \begin{tabular}{c c c}
  \hline
    CaseID & Activity & Actor    \\ \hline
1      & A        & Matt     \\
2      & A        & Matt     \\
1      & B        & Britney  \\
1      & E        & Matt     \\
2      & E        & Matt     \\
2      & B        & Britney \\
3      & A        & Brad     \\
3      & E        & Matt     \\
4      & A        & Brad     \\
5      & A        & Brad     \\
3      & B        & Brad     \\
4      & B        & Britney  \\
4      & E        & Brad     \\
6      & A        & Brad     \\
5      & B        & Joan     \\
6      & C        & Joan     \\
5      & E        & Brad     \\
1      & D        & George   \\
6      & D        & George  \\ \hline
  \end{tabular}
  }
  {\caption{\small Event Log}}
 \ttabbox
  { \begin{tabular}{|c|c|c|c|c|c|}
\hline
        & A & B & C & D & E \\ \hline
Matt    & 2 & 0 & 0 & 0 & 3 \\ \hline
Britney & 0 & 3 & 0 & 0 & 1 \\ \hline
Brad    & 4 & 1 & 0 & 0 & 1 \\ \hline
Joan    & 0 & 1 & 1 & 0 & 0 \\ \hline
George  & 0 & 0 & 0 & 2 & 0 \\ \hline
\end{tabular}
  }
 {\caption{\small Actor-Activity Matrix}}
\end{floatrow}
\end{table}
\subsection{Similar-Task Algorithm}
\par{Similar-Task algorithm which comes under Organizational Mining is a metric based on joint activities. It does not consider how individuals work together on shared cases but focuses on the activities they perform \cite{aalst}. Similar-Task aims at finding similarity between Actors based on the intersection of Activities. The idea is that individuals performing similar tasks are more closely related to each other than individuals performing different tasks  \cite{aalst}. Similarity calculation could be achieved using Cosine-Similarity, Pearson Correlation Coefficient, Hamming Distance, etc. Based on previous literature reviews, we present the following adaptation of Similar-Task algorithm.}
\par{The input to Similar-Task algorithm is a 2-dimensional matrix. The matrix contains frequencies of activities performed by each actor. This matrix is commonly referred as Actor-Activity Matrix. An example of Actor-Activity Matrix is shown in Table 2. For instance, Matt performs activity A twice, activity E thrice and has no involvement in activities B, C and D. In this paper, we use Cosine-Similarity as a metric of measuring similarity between Actors based on the Activities they perform. Table 3 gives similarity values between Actors based on Algorithm 1.}
\begin{algorithm}[t]
\KwData{Actor-Activity Matrix (M)}
\KwResult{Matrix with similarity values between Actors}
Get the number of rows of M into \textbf{$m$}.\\
Get the number of columns of M into \textbf{$n$}.\\
$D[m][m]$ = Declare square matrix to store results.\\
\ForEach { \textnormal{i = $1$ to $m-1$ } }{
$P$=Vector corresponding to $i^\text{th}$ row.\\
\ForEach { \textnormal{j = $i+1$ to $m$ } }{
$Q$=Vector corresponding to $j^\text{th}$ row.\\
Apply Cosine Similarity between $i^\text{th}$ and $j^\text{th}$ row
\begin{equation}
\cos(P,Q) = \frac{P \cdot Q}{\parallel P \parallel \parallel Q \parallel}
\end{equation}\\
Set $D[i][j]$=similarity value obtained in the Step 8.\\
}
}
\caption{Similar-Task Algorithm}
\label{alg:Similar-Task Algorithm}
\end{algorithm}

\subsection{Subcontract Algorithm}
\par{Sub-Contract is another Organizational Mining metric  which is based on causal dependencies between Actors in carrying out business process \cite{aalst}. Sub-Contract Algorithms tries to find out the number of times individual $j$ executes it's task in between two activities performed by individual $i$ \cite{aalst}. Sub-Contract algorithm considers dependencies between activities in the process model, commonly referred as \textit{causality fall factor} ($\beta$). These dependencies can be obtained using Process Discovery techniques like $\alpha$-miner algorithm. Sub-Contract algorithm also considers direct/indirect succession ($depth$) between Actors. It also takes into consideration whether sub-contraction between Actors occurs single or multiple times. Sub-Contract algorithm presented in Algorithm 2 considers indirect succession, multiplicity while ignoring dependencies of activities.}
\par{Each $Process Instance$ corresponds to a Case Identifier (CaseID) in the event log. \textit{AuditTrailEntryList} constitutes all the events pertaining to a particular CaseID. An $AuditTrailEntry$ refers to an individual event \cite{mxml}. For example, considering events pertaining to Case1 in Table 1, we have a sub-contraction between Matt and Britney. Matrix entry corresponding to Matt and Britney is updated in $m$ followed by an update in $D$. Final result shown in Table 4 is obtained after all such sub-contractions are identified from all cases in the event log.}
\renewcommand{\arraystretch}{.5}
\begin{table}
\CenterFloatBoxes
\begin{floatrow}
\ttabbox{
 \begin{tabular}{|l|l|l|l|l|l|}
\hline
        & Matt                    & Britney                 & Brad                    & Joan                    & George                  \\ \hline
Matt    & \multicolumn{1}{c|}{--} & 0.263                   & 0.719                   & 0.00                    & 0.00                    \\ \hline
Britney & \multicolumn{1}{c|}{--}                   & \multicolumn{1}{c|}{--} & 0.298                   & 0.671                   & 0.00                    \\ \hline
Brad    & \multicolumn{1}{c|}{--}                   & \multicolumn{1}{c|}{--}                   & \multicolumn{1}{c|}{--} & 0.167                   & 0.00                    \\ \hline
Joan    & \multicolumn{1}{c|}{--}                    & \multicolumn{1}{c|}{--}                   & \multicolumn{1}{c|}{--}                   & \multicolumn{1}{c|}{--} & 0.00                    \\ \hline
George  & \multicolumn{1}{c|}{--}                    & \multicolumn{1}{c|}{--}                    & \multicolumn{1}{c|}{--}                    & \multicolumn{1}{c|}{--}                    & \multicolumn{1}{c|}{--} \\ \hline
\end{tabular}
 }
 {\caption{\small Cosine-Similarity Values} }
\ttabbox
  {
\begin{tabular}{|c|c|c|c|c|c|}
\hline
        & Matt & Britney & George & Brad & Joan \\ \hline
Matt    & 0    & 0.22    & 0      & 0    & 0    \\ \hline
Britney & 0    & 0       & 0      & 0    & 0    \\ \hline
George  & 0.22 & 0       & 0      & 0    & 0    \\ \hline
Brad    & 0    & 0.22    & 0      & 0    & 0.22 \\ \hline
Joan    & 0    & 0       & 0      & 0    & 0    \\ \hline
\end{tabular}
}
{\caption{Sub-contract values }}
\end{floatrow}
\end{table}

\begin{algorithm}
\KwData{$\beta$, $depth$, $Len$, $Log$}
\KwResult{Normalized 2D Matrix $D$ with subcontract values between Actors}
Declare Square Matrix $D$ of size $Len$*$Len$. Initialize all elements to 0\\
Declare and initialize variable $normal$ to 0\\
\ForEach { \textnormal{ProcessInstance} $pi$ \textnormal {in the log}}{
Get $AuditTrailEntryList$ $ates$ for $pi$\\
\If { \textnormal{ $size_{ates}$ $< 3$  }}{
  continue to the next ProcessInstance, $pi$\\
  }
  Declare and intialize $minK$ to 0.\\
 \If{$size_{ates}$ $<$ $depth$}{
 set $minK$= $size_{ates}$\\
 }
\Else{ set $minK$=$depth$ + 1. \\ }
\If{$minK$ $<$ 3}{
set $minK$=3.}
\ForEach {\textnormal{k:=$2$ to $minK$}}{
Update $normal$ by $\beta^\textnormal{k-2}$.\\
$m$= Square matrix of $Len$*$Len$.\\
\ForEach {\textnormal{i:=$0$ to $size_{ates}-k$}}{
$ate_i$ = get $AuditTrailEntry$ at position $i$.\\
$ate_{ik}$ = get $AuditTrailEntry$ at position $i+k$\\
\If{$Actor_{ate_i}$ = $Actor_{ate_{ik}}$}{
\ForEach {\textnormal{j:=$i+1$ to $i+k$}}{
$ate_j$ = get $AuditTrailEntry$ at position $j$.\\
$row$ = get row-position for $Actor_{ate_i}$\\
$col$ = get column-position for $Actor_{ate_j}$\\
For valid ($row$ , $col$ ) set m[$row$][$col$]=1.
}
}
}
\ForEach {\textnormal{i:=$0$ to $Len$}}{
\ForEach {\textnormal{j:=$0$ to $Len$}}{
set D[$i$][$j$] = D[$i$][$j$] + m[$i$][$j$]*$\beta^\textnormal{k-2}$.
}
}
}
}
Return $Normalized Matrix D$. \textnormal{//divide each value by $normal$.}\\
\caption{Sub-Contract Algorithm }
\label{alg:Subcontract Algorithm}
\end{algorithm}
\begin{algorithm}
\KwData{Actor-Activity Graph}
\KwResult{Graph with similarity values between Actors}
$A_{i}$ = Get an Actor 'i' from the Actor-Activity Graph.\\
$A_{j}$ = Get another Actor 'j' from the Actor-Activity Graph.\\
Find intersecting Activities between $A_{i}$ and $A_{j}$.\\
Collect frequencies of Activities from the edges of intersecting Activities.\\
Apply Cosine-Similarity with the values obtained in Step 4.\\
Set [:SIMILARITY] between $A_{i}$ and $A_{j}$ with the value obtained in Step 5.
\caption{Similar-Task Algorithm in Graph Database}
\end{algorithm}

%

\section{\textbf{Implementation of Algorithms on RDBMS}}
We present a few segments of our implementation due to limited space in the paper. The entire code and implementation can be downloaded from our website\footnote{http://goo.gl/wMyUOS}.
\subsection{Similar-Task Algorithm}
\par{Typical Steps involved in the implementation of Similar-Task algorithm in RDBMS is shown in Fig. 1. We import the event log dataset into a table, \textit{dataset}. A stored procedure creates Actor-Activity matrix (a table in MySQL) from dataset table. We use Actor-Activity matrix(AAMatrix) for calculating cosine-similarity in another stored procedure and the similarity values are collected in Result Matrix.}
\begin{figure}
  \centering
  \caption{Similar-Task implementation flow in RDBMS}
  \includegraphics[width=\linewidth,height=2.5cm]{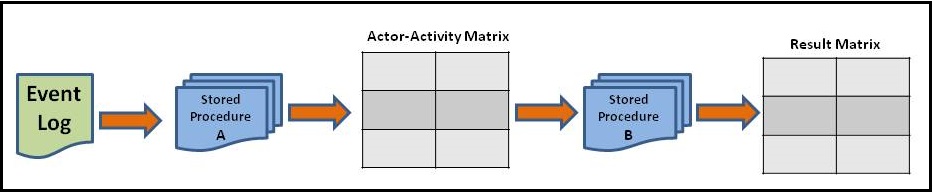}
 \end{figure}
\par{The SQL implementation of Similar-Task algorithm involves Create, Read, Update, Delete (CRUD)\footnote{http://dev.mysql.com/doc} statements. We define these statements as a single adhoc SQL query or as part of stored procedure(s).
\begin{enumerate}[nolistsep]
\item To create Actor-Activity matrix, we define a stored procedure that takes the table dataset as input parameter. 
    \begin{enumerate}
    \item We collect all distinct Activity from table dataset using a cursor\footnote{http://dev.mysql.com/doc/refman/5.0/en/cursors.html}.
    \item We create a table \textit{AAMatrix} to store frequency of each Activity performed by the Actors. AAMatrix's schema is of the form (Actor, Activity1, Activity2,...) where Actor is of type VARCHAR and is a PRIMARY KEY, and Activity1, Activity2, etc. are all those Activities collected from the cursor and are of type INT.
    \end{enumerate}
\item We populate Actor-Activity matrix using INSERT and IF statements inside the stored procedure. For any (Actor, Activity) pair that is found, its corresponding value in AAMatrix is incremented by one.
\lstset{frame=tb,                       
  language=SQL,
  aboveskip=2mm,
  belowskip=2mm,
  showstringspaces=false,
  columns=flexible,
  basicstyle={\small\ttfamily},
  numbers=none,
  numberstyle=\tiny\color{gray},
  keywordstyle=\color{blue},
  commentstyle=\color{dkgreen},
  stringstyle=\color{mauve},
  breaklines=true,
  breakatwhitespace=true,
  tabsize=3
}
\lstset{language=SQL} 
\begin{lstlisting}{create}
COUNT(IF (ACTIVITY='ACTIVITY1', 1, NULL))
COUNT(IF (ACTIVITY='ACTIVITY2', 1, NULL))
\end{lstlisting}
This combination of COUNT and IF statements are combined with INSERT statement to populate AAMatrix.
\item Calculation of Cosine-Similarity is done using another stored procedure that takes Actor-Activity Matrix as input parameter. A table \textit{InitSim} with schema (SOURCEACTOR, TARGETACTOR, SIMILARITY) is created to store similarity values as they are calculated. SOURCEACTOR and TARGETACTOR are of type VARCHAR, while SIMILARITY is of type DOUBLE. Join is applied to two instances of AAMatrix and cosine-similarity calculated for each pair of distinct Actors.
\lstset{frame=tb,                       
  language=SQL,
  aboveskip=2mm,
  belowskip=2mm,
  showstringspaces=false,
  columns=flexible,
  basicstyle={\small\ttfamily},
  numbers=none,
  numberstyle=\tiny\color{gray},
  keywordstyle=\color{blue},
  commentstyle=\color{dkgreen},
  stringstyle=\color{mauve},
  breaklines=true,
  breakatwhitespace=true,
  tabsize=3
}
\lstset{language=SQL} 
\begin{lstlisting}{create}
AAMatrix T1 JOIN AAMatrix T2
WHERE T1.ACTOR <> T2.ACTOR
\end{lstlisting}
The similarity values obtained with calculations on the join are first ordered by T1.ACTOR, followed by T2.ACTOR and then inserted into InitSim.
\item We create another table \textit{FinalSim} with schema (SOURCEACTOR, ACTOR1, ACTOR2, ...) and populate it using values from InitSim. The schema is also created using cursors in the stored procedure where each distinct Actor forms the column of table FinalSim. Data into table FinalSim is populated in the same way as Step 2.
\end{enumerate}
\subsection{Sub-Contract Algorithm}
\par{Alike Similar-Task algorithm, implementation of Sub-Contract algorithm also involves ad-hoc SQL queries and dynamically built queries using stored procedures. The implementation flow is much alike Fig. 1, and has not been shown here.
\begin{enumerate}
\item We import the event log dataset in a table also named \textit{dataset} (ID, CaseID, Actor, Activity) where ID is an auto incrementing field, CaseID is the case identifier selected from the dataset. Actor and Activity are self-explanatory. However, for efficient implementation of the algorithm, data from the \textit{dataset} is re-ordered so that all events corresponding to a CaseID are together and ordered in ascending order. We define a secondary table named \textit{organiseddata} to store this ordered information.
\item Sub-Contraction can only be detected if there are at least three (3) events in a particular CaseID. Joins are applied only when this criteria is met. Since each event in the table is assigned a unique ID (auto incrementing), so actor responsible for sub-contraction will always have ID difference of at least 2. The following SQL snippet joins tables for each CaseID to find the IDs of actors responsible for sub-contraction.
\lstset{frame=tb,                       
  language=SQL,
  aboveskip=2mm,
  belowskip=2mm,
  showstringspaces=false,
  columns=flexible,
  basicstyle={\small\ttfamily},
  numbers=none,
  numberstyle=\tiny\color{gray},
  keywordstyle=\color{blue},
  commentstyle=\color{dkgreen},
  stringstyle=\color{mauve},
  breaklines=true,
  breakatwhitespace=true,
  tabsize=3
}
\lstset{language=SQL} 
\begin{lstlisting}{create}
    organiseddata AS T1 JOIN organiseddata AS T2
    ON T2.ID >= T1.ID + 2
    AND T1.ACTOR= T2.ACTOR 
    AND T1.ACTIVITY <> T2.ACTIVITY
    ORDER BY DIFF ASC
\end{lstlisting}
\item Once IDs of Actors responsible for sub-contraction are found out in Step 2, all intermediate IDs are collected and their sub-contraction strength calculated. We create a table \textit{RESULTTABLE} (PERFORMER, ACTOR1, ACTOR2, ...) to store the value of sub-contraction. Here, PERFORMER is the actor with whom sub-contraction is being considered, and ACTOR1, ACTOR2, etc. are other Actors that are placed dynamically using stored procedure.
\end{enumerate}
%

\section{\textbf{Implementation of Algorithms in Neo4j}}
\subsection{Similar-Task Algorithm}
\par{The Steps involved in the implementation of Similar-Task algorithm in Neo4j\footnote{http://neo4j.com/docs/} is shown in Fig. 2. Fig. 2(a) depicts how Actor-Activity information is maintained in Neo4j. While Fig. 4(b) presents a typical view of the graph after similarity calculation. We present Similar-Task algorithm adapted for graph database in Algorithm 3.}
\begin{figure}
  \centering
  \mbox{
    \subfigure[Actor-Activity information in Graph Database]{\includegraphics[width=9cm,height=4.5cm]{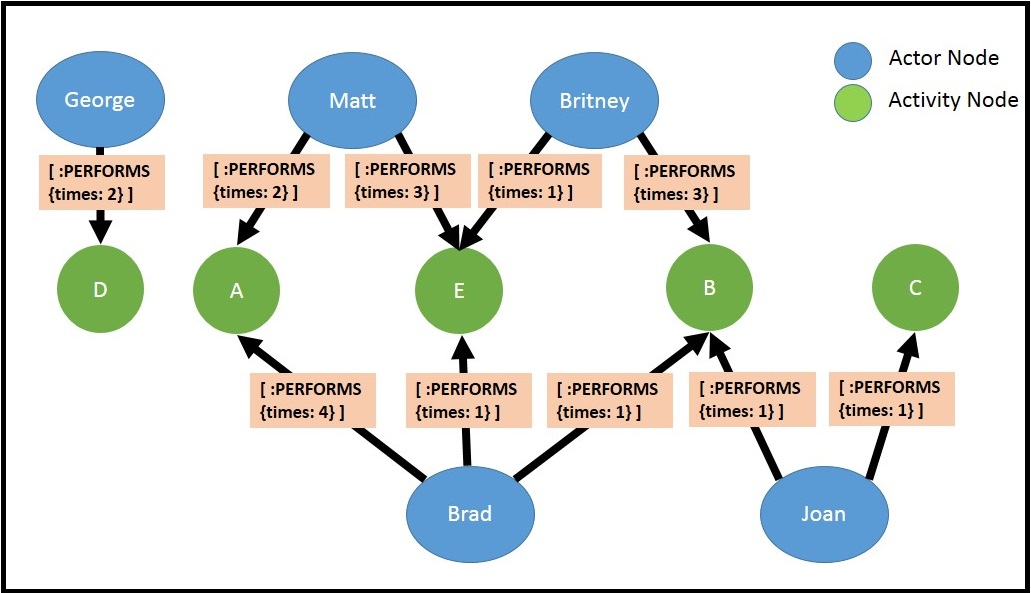}}\quad
    \subfigure[Similarity values in Graph Databases]{\includegraphics[width=6.5cm,height=4.5cm]{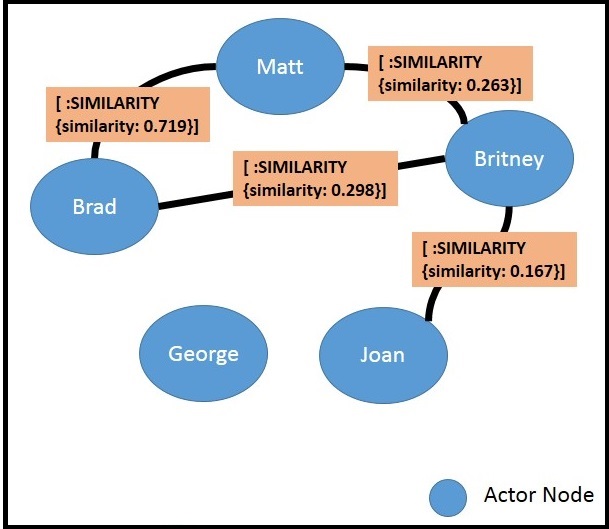}}
  }
  \caption{Similar-Task implementation flow in Graph Database}
\end{figure}
\par{\begin{enumerate}[nolistsep]
\item We create nodes and relationships such that Actor-Activity information is also calculated and stored directly during dataset import. We create only unique Actor and Activity nodes and merge the relationship between them for any repetition. Relationship [:PERFORMS] connects an Actor to an Activity node with a property \textit{times} that records the frequency of the Activity performed by that Actor.
\item Calculation of Cosine-Similarity in Neo4j comprises of three steps.
\begin{enumerate}
\item All intersecting Activities between a pair of Actors are found out. 
\item Using value of \textit{times} property from [:PERFORMS] relationship for all the intersecting activities found in Step 2(a), cosine-similarity is calculated. 
\item The cosine-similarity value thus obtained is stored as a value to property \textit{similarity} in the relationship [:SIMILARITY] between the Actors in consideration.
\lstset{frame=tb,                       
  language=SQL,
  aboveskip=1mm,
  belowskip=1mm,
  showstringspaces=false,
  columns=flexible,
  basicstyle={\small\ttfamily},
  numbers=none,
  numberstyle=\tiny\color{gray},
  keywordstyle=\color{blue},
  commentstyle=\color{dkgreen},
  stringstyle=\color{mauve},
  breaklines=true,
  breakatwhitespace=true,
  tabsize=3
}
\lstset{language=SQL} 
\begin{lstlisting}{create}
MATCH (p1:Actor)-[x:PERFORMS]->(m:Activity)<-[y:PERFORMS]-(p2:Actor) 
WITH SUM(x.times * y.times) AS xyDotProduct, 
	SQRT(REDUCE(xDot = 0.0, a IN COLLECT(x.times) | xDot + a^2)) AS xLength, 
	SQRT(REDUCE(yDot = 0.0, b IN COLLECT(y.times) | yDot + b^2)) AS yLength, 
	p1, p2 
MERGE (p1)-[s:SIMILARITY]-(p2) 
SET s.similarity = xyDotProduct / (xLength * yLength)
\end{lstlisting}
\end{enumerate}
\end{enumerate}
}
\subsection{Sub-Contract Algorithm}
\par{We implement Sub-Contract algorithm in graph database using a similar approach as shown in Fig. 2. But unlike Similar-Task algorithm implementation in CYPHER, only CaseID is made unique. Whereas other information like Actor and Activity are created for each event in the event log.}
\par{\begin{enumerate}
\item We create unique 'CASE' nodes for distinct CaseIDs. These Case nodes stores information like the case names, and an incrementing counter, occurrence ID (OccID) whose value increases as new Actor nodes for that CaseID is added. Within each CaseID, Actor nodes are created with information like actor name, OccID (taken from CaseID) and the activity it performs. Case nodes and Actor nodes are connected via [:CONTAINS] relationship.

\item Second Step involves finding Actors responsible for sub-contraction. It is worth mentioning that OccID are always assigned in ascending order and only those Actors with same name but with different Activity and OccID are responsible for sub-contraction. For each CaseID,
\begin{enumerate}
\item We find out OccIDs of the Actors responsible for sub-contraction.
\item We collect all intermediate OccIDs between the OccIDs found in Step 2(a).
\item A relationship [:RELATED\textunderscore TO] is created from the the OccID of starting Actor node found in Step 2(a) to all intermediate OccIDs found in Step 2(b). A property \textit{value} is set to 1 that is used in subsequent steps.
\end{enumerate}
\lstset{frame=tb,                       
  language=SQL,
  aboveskip=2mm,
  belowskip=2mm,
  showstringspaces=false,
  columns=flexible,
  basicstyle={\small\ttfamily},
  numbers=none,
  numberstyle=\tiny\color{gray},
  keywordstyle=\color{blue},
  commentstyle=\color{black},
  stringstyle=\color{mauve},
  breaklines=true,
  breakatwhitespace=true,
  tabsize=3
}
\lstset{language=SQL} 
\begin{lstlisting}{create}
WITH commActorPath, n, (Actor2.OccID - Actor1.OccID) as sepDist
WITH RANGE(head(nodes(commActorPath)).OccID+1, last(nodes(commActorPath)).OccID-1) as intermediateIDs,
     n, head(nodes(commActorPath)).OccID as startID, sepDist
UNWIND intermediateIDs as endID
MATCH (person1:PERSON {OccID:startID})<--(n)-->(person2:PERSON {OccID:endID})
MERGE (person1)-[:RELATED_TO {value:1, length:sepDist}]->(person2)
\end{lstlisting}
We define commActorPath as the path to find out the Actors (with common name) responsible for sub-contraction. RANGE function collects all the intermediate IDs which is then used to connect sub-contracting Actors.
\item Sub-contraction strength between two Actors can only be set once they Actor nodes are made unique. To do so, we create UNIQUEACTOR nodes for all distinct Actor names in the database.
\item The final Step of the algorithm involves setting sub-contract strength between UNIQUEACTOR nodes. For each CaseID,
\begin{enumerate}
\item We collect start node and end node of [:RELATED\textunderscore TO] relationship.
\item We refer the UNIQUEACTOR nodes corresponding to the start node and end node found in Step 4(a).
\item We establish sub-contraction between the two UNIQUEACTOR nodes using [:SUBCONTRACT] relationship with a property \textit{strength} whose value is updated accordingly using the \textit{value} property of [:RELATED\textunderscore TO] relationship.
\end{enumerate}
\lstset{frame=tb,                       
  language=SQL,
  aboveskip=2mm,
  belowskip=2mm,
  showstringspaces=false,
  columns=flexible,
  basicstyle={\small\ttfamily},
  numbers=none,
  numberstyle=\tiny\color{gray},
  keywordstyle=\color{blue},
  commentstyle=\color{dkgreen},
  stringstyle=\color{mauve},
  breaklines=true,
  breakatwhitespace=true,
  tabsize=3
}
\lstset{language=SQL} 
\begin{lstlisting}{create}
MATCH (n)-[:CONTAINS]->()-[r:RELATED_TO]->()<-[:CONTAINS]-(n)
MERGE (p:UNIQUEACTOR {name:startNode(r).name})-[rf:SUBCONTRACT]->(q:UNIQUEACTOR {name:endNode(r).name})
SET rf.strength = CASE WHEN rf.strength IS NULL THEN r.value ELSE rf.strength + (0.5^(l-2))*r.value END
\end{lstlisting}
\end{enumerate}
}

%

\section{\textbf{Experimental Dataset}}

\par{We conduct experiments on a publicly available large real world dataset downloaded from Business Process Intelligence $2014$ (BPI $2014$)\footnote{http://www.win.tue.nl/bpi/2014/start}. The dataset contains information on Information Technology Infrastructure Library (ITIL) of Robobank Information and Communication and Technology (ICT). ITIL is a process of addressing customer grievances regarding disruption in ICT services. A Service Desk Agent records the complete information about the problem in an Interaction record.  We choose the 'Detail Incident Activity' for our set of experiments. The dataset contains $4,66,737$ records and out of the seven fields  in the dataset, we use the following three}
\par{\begin{enumerate}[nolistsep]
\item Incident\textunderscore ID: The unique ID of a record in the Service Management tool. It is represented as CaseID in our data model.
\item IncidentActivity\textunderscore Type: Identifies which type of an activity takes place.
\item Assignment\textunderscore Group: The team responsible for an activity.
\end{enumerate}
}

%

\section{Benchmarking and Performance Comparison}
\par{We conduct a series of experiments on the implementations of Similar-Task and Sub-Contract algorithms. Our benchmarking system consists of Intel Core 2 Duo (3M Cache, 2.1 GHz), 4 GB DDR3 RAM and 320 GB of Hard disk drive. We use Windows 8.1 with single node setup of MySQL 5.6 and Neo4j 2.14. We ensure that only minimally required services are running during the analysis. We conduct experiments on warmed up cache and the values recorded are an average over five runs of the implementations. In order to study scalability, we divide our event log dataset into five different chunks of increasing size and conduct experiments that takes into consideration both the size of the dataset and the number of unique actors in each chunk. Table 5 presents the statistics for each of these chunks.}
\subsection{Similar-Task Algorithm}
\par{Table 6 and Fig. 3 reveals load time across different sizes. The load time includes loading data into a table, processing it to generate and populate an Actor-Activity matrix. The load time varies with the number of unique actors present in each chunk. We observe that both the databases give similar performance. However with increase in number of unique actors, Neo4j gives better load time performance and has been seen to perform 1.25x magnitude faster than MySQL.}
\renewcommand{\arraystretch}{.5}
\begin{table}
\CenterFloatBoxes
\begin{floatrow}
\ttabbox
  { \begin{tabular}{|c|c|}
  \hline
       \textbf{Dataset Size} & \textbf{Unique Actors}\\
    \hline
     65000 & 150 \\ \hline
     1,01,000 & 158 \\ \hline
     2,19,500 & 220 \\ \hline
     3,00,000 & 229 \\ \hline
     4,66,737 & 242  \\ \hline
  \end{tabular}
  }
{\caption{\small \specialcell[c]{Number of Unique\\Actors per dataset size.}}}
 \ttabbox
  { \begin{tabular}{|c|c|c|}
  \hline
       \textbf{Unique Actors} & \multicolumn{2}{|c|}{ \textbf{Load Time (msec)}}\\
    \hline
    & \textbf{MySQL}& \textbf{Neo4j}\\ \hline
     150 & 2467 &  3413    \\ \hline
     158 & 2875 & 3362   \\ \hline
     220 & 5966 & 4354    \\ \hline
     229 & 5850 & 5877    \\ \hline
     242 & 7819 & 6875    \\ \hline
  \end{tabular}
  }
{ \caption{\small \specialcell[c]{Data Load Time\\($Similar-Task$)}}}
\end{floatrow}
\end{table}

\begin{figure}
\centering
\includegraphics[width=9cm,height=5cm]{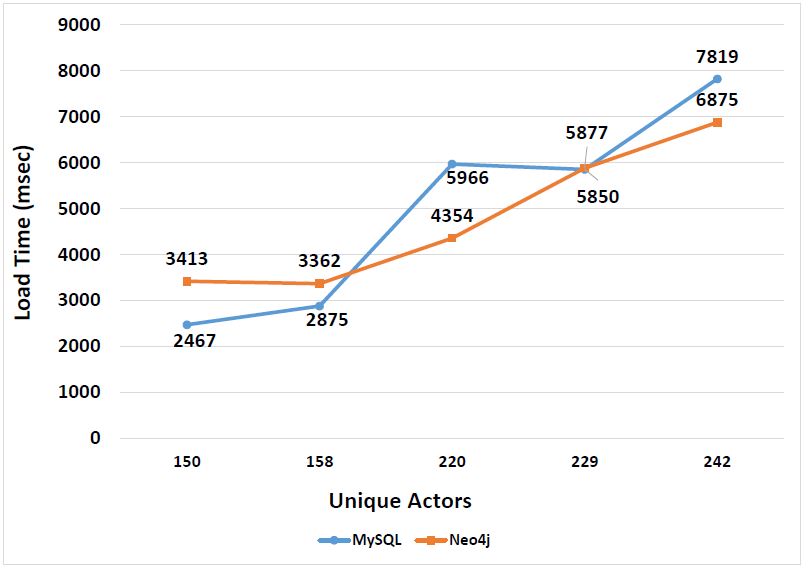}
\caption{Data Load Time for Similar-Task Algorithm}
\end{figure}
\par{We believe that Neo4j's better performance is due to the fact that only unique Actor and Activity nodes are created during data import. On the other hand, MySQL has a predefined schema with number of columns being equivalent to the number of distinct activities in the dataset. Hence even if an actor has not performed an activity, value (albeit zero or default) has to be set at that respective column. Whereas Neo4j defines relationship only when they are discovered and thus gives better performance.}

\par{The core of Similar-Task algorithm is similarity calculation between Actors (Step-8 of the Algorithm 1) and updating the result table (Step-9 of the Algorithm 1). Table 7 and Fig. 4 displays the time taken to calculate cosine-similarity and update result in Similar-Task Algorithm as a function of the number of unique actors for different dataset chunk (given in Table 5). It is interesting to note that execution time for cosine-similarity calculation in MySQL is 32 times better than Neo4j. In case of write operations too, MySQL slightly outperforms Neo4j by a magnitude of 1.5.}
\renewcommand{\arraystretch}{.5}
\begin{table}[H]
\centering
\caption{\small Execution Time for Step-8 and Step-9 ($Similar-Task$)}
\begin{tabular}{|c|c|c|c|c|}
  \hline
      \textbf{Unique Actors} & \multicolumn{4}{|c|}{\textbf{Execution Time (msec)}}\\
    \hline
    & \multicolumn{2}{|c|}{\textbf{Step-8}} & \multicolumn{2}{|c|}{\textbf{Step-9}}\\ \hline
    & \textbf{MySQL}& \textbf{Neo4j} & \textbf{MySQL}& \textbf{Neo4j}\\ \hline
     150 & 225 & 9616 & 1907 & 2403\\ \hline
     158 & 372 & 11700 & 2844 & 2925\\ \hline
     220 & 713 & 14655 & 6292 & 3664\\ \hline
     229 & 903 & 29520 & 6703 & 7380\\ \hline
     242 & 1403 & 48891 & 8453 & 12223\\ \hline
  \end{tabular}
\end{table}


\begin{figure}
  \centering
  \mbox{
    \subfigure[Execution time for cosine-similarity calculation]{\includegraphics[width=7.5cm,height=5cm]{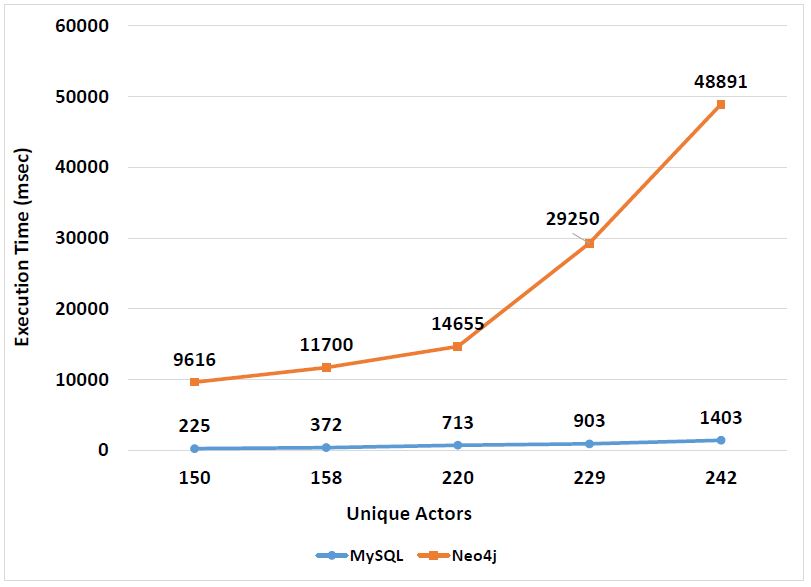}}\quad
    \subfigure[Time taken to update results]{\includegraphics[width=7.5cm,height=5cm]{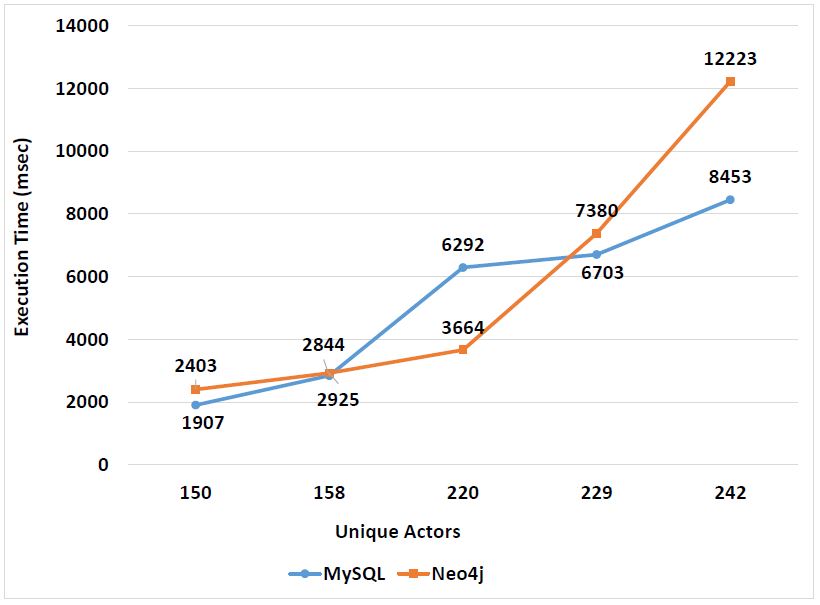}}}
  \caption{Execution Time for Step-8 and Step-9 (Similar-Task)}
\end{figure}
\par{We believe that data meant for cosine similarity calculation in MySQL is available in tables, and fetching these data is then only limited to advancing pointers to the next row. Graph databases like Neo4j are not known to have such constructs available for matrix. Calculations in Neo4j are based on reading property values defined on nodes and relationships. Cosine-similarity computation in Neo4j requires matching intersecting activities between the two actors in concern, collecting property values from the relationships connecting these intersecting activities, followed by the actual computation. It is for this reason that we see a sharp rise in cosine-similarity calculation in Neo4j. Fig. 4(b) gives an estimate of the time required to update results. We observe that setting properties in Neo4j is more time consuming because existing relationships needs to be merged with the updated property values or new ones be created if such relationship does not exist. Whereas updating results in MySQL only consists of updating results in respective columns on an already defined table.}

\renewcommand{\arraystretch}{.5}
\begin{table}[H]
\CenterFloatBoxes
\begin{floatrow}
\ttabbox
  {\tabcolsep=0.06cm 
  \begin{tabular}{|c|c|c|c|c|c|}
  \hline
      \textbf{Tables} & \multicolumn{5}{|c|}{\textbf{Dataset Size}}\\
    \hline
    & \textbf{65000}& \textbf{101000} & \textbf{219500}& \textbf{300000} & \textbf{466737}\\ \hline
     Dataset & 3686400 & 5783552 & 11026432 & 15220736 & 21544960\\ \hline
    AAMatrix & 65536 & 65536 & 65536 & 81920 & 81920\\ \hline
    InitSim & 1589248 & 1589248 & 1589248 & 3686400 & 3686400\\ \hline
    FinalSim & 229376 & 262144 & 278528 & 491520 & 1589248\\ \hline
  \end{tabular}
  }
{\caption{\small \specialcell[c]{Disk Space Usage (bytes) for\\MySQL tables ($Similar-Task$)}}}
 \ttabbox
  {\tabcolsep=0.06cm 
  \begin{tabular}{|c|c|c|c|c|c|}
  \hline
       \textbf{\specialcell[c]{Graph\\Elements}} & \multicolumn{5}{|c|}{ \textbf{Dataset Size}}\\
    \hline
    & \textbf{65000}& \textbf{101000}& \textbf{219500} & \textbf{300000} & \textbf{466737}\\ \hline
     Nodes & 2820 & 2910 & 3075 & 3990 & 4215\\ \hline
Relationships & 770040 & 414315 & 479663 & 856809 & 983227\\ \hline
Properties & 1033856 & 563873 & 651203 & 1155011 & 1323439\\ \hline

\end{tabular}
  }
{ \caption{\small \specialcell[c]{Disk Space Usage (bytes) for\\Neo4j Elements ($Similar-Task$)}}}
\end{floatrow}
\end{table}


\par{Table 8 and Fig. 5(a) presents the disk space taken by tables in MySQL which includes both the space taken by actual data and indexes, if any. Readers are suggested to refer to Section 4.1 for better understanding of the tables associated with the implementation of Similar-Task algorithm. Table 9 and Fig. 5(b) shows disk space taken by various graph elements in Neo4j. We observe that Neo4j uses almost 12 times less disk space in comparison to MySQL. We believe that nodes and relationships in Neo4j are created only when needed. On the other hand, MySQL needs to write values for all columns which contributes to higher disk usage.}
\begin{figure}
  \centering
  \mbox{
    \subfigure[Disk Space Usage for MySQL tables]{\includegraphics[width=7.5cm,height=5cm]{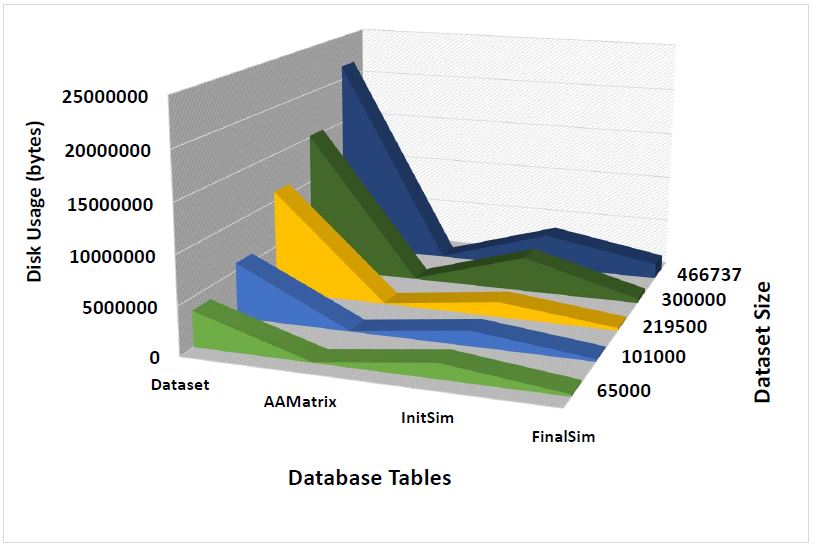}}\quad
    \subfigure[Disk Space Usage for Neo4j elements]{\includegraphics[width=7.5cm,height=5cm]{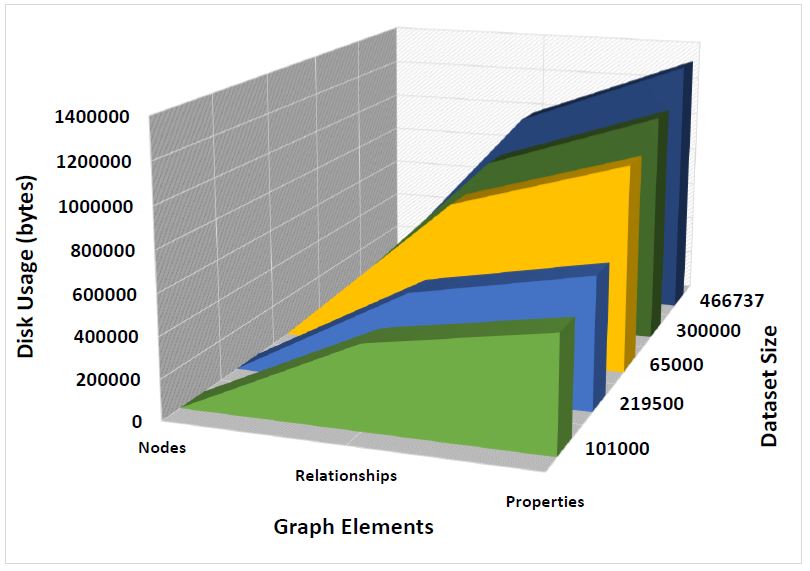}}}
  \caption{Disk Space Usage in Similar-Task algorithm.}
\end{figure}
\subsection{Sub Contract Algorithm}
\par{Table 10 and Fig. 6 shows data load time across different dataset sizes. The load time includes loading the event log dataset, pre-procesing and writing it back to the database. Pre-procesing in MySQL involves ordering the event log dataset by CaseID, whereas assigning incremental occurrence identifiers to Actor nodes within each Case node in Neo4j. We observe that for a single node setup, both the databases give similar performance. However with increase in dataset size, Neo4j gives better load time performance and is seen to perform 1.15x magnitude faster than MySQL. Also data load time in Sub-Contract algorithm is 5.5x magnitude slower than Similar-Task algorithm.}

\renewcommand{\arraystretch}{.5}
\begin{figure}[H]
\CenterFloatBoxes
\begin{floatrow}
\ttabbox
  {\begin{tabular}{|P{1.4cm}|P{1.3cm}|P{1.2cm}|}
  \hline
       \textbf{\specialcell[c]{DataSet\\Size}} & \multicolumn{2}{|c|}{ \textbf{\specialcell[c]{Load Time\\(msec)}}}\\
    \hline
    & \textbf{MySQL}& \textbf{Neo4j}\\ \hline
     65,000 & 6575 &  9567    \\ \hline
     1,01,000 & 8390 & 10476   \\ \hline
     2,19,500 & 14279 & 14873    \\ \hline
     3,00,000 & 26437 & 25435    \\ \hline
     4,66,738 & 43712 & 38234    \\ \hline
  \end{tabular}
  }
  {\caption{\small \specialcell[c]{Data Load Time\\($Sub-Contract$)}}}
  \killfloatstyle
\ffigbox
{\caption{\specialcell[c]{Data Load Time for\\Sub-Contract Algorithm}}}
  {\includegraphics[width=9cm,height=5cm]{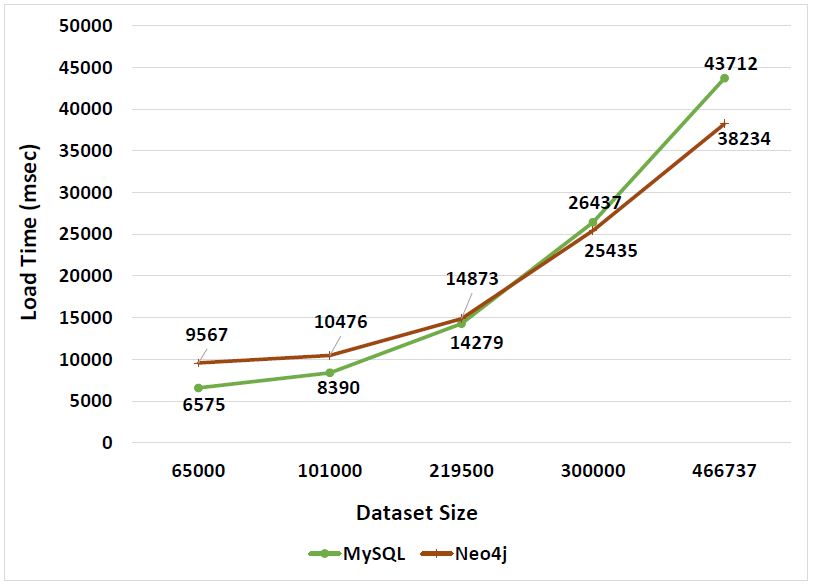}}
\end{floatrow}
\end{figure}
\par{We observe that alike Similar-Task algorithm, data load time exhibits similar pattern in Sub-Contract algorithm too. With increase in dataset size, ordering tables by CaseID and writing them back to database takes longer time as compared to creating nodes in Neo4j. However we observe that load time is higher in Sub-Contract algorithm as compared to Similar-Task algorithm because Actor nodes are created for each event in the event log. Whereas in Similar-Task algorithm only unique nodes are created. We believe that setting property values on nodes for each event in the dataset incurs more write operations and thus takes more time as compared to setting property values for unique nodes in Similar-Task algorithm.}

\par{Table 11 displays execution time of four major steps of Sub-Contract algorithm. These steps include updating the value of normal (Update Normal), detecting sub-contracting Actors (Sub-Contraction Detection), writing the result back to the database (Update Result) and normalizing the result (Normalize Result). Table 12 shows the execution time noted for four major steps of Sub-Contract algorithm implemented in Neo4j. We record execution time for the four major steps as a function of dataset size and the results are presented in Fig. 7(a) and Fig. 7(b). We observe that Sub-Contract algorithm implemented in MySQL have identical performance across dataset chunks. On the other hand, Sub-Contract algorithm's performance in Neo4j varies linearly with increase in dataset size. We observe that detecting sub-contracting Actors in Neo4j attains performance boost of the magnitude of 7x over MySQL. Empirical analysis shows that write operations in MySQL is almost 4 times slower than Neo4j.}

\renewcommand{\arraystretch}{.5}
\begin{table}
\CenterFloatBoxes
\begin{floatrow}
\ttabbox
  {\tabcolsep=0.06cm 
  \begin{tabular}{|c|c|c|c|c|}
  \hline
      \textbf{\specialcell[c]{Dataset\\Size}} & \multicolumn{4}{|c|}{\textbf{Execution Time(msec)}}\\
    \hline
    & \textbf{\specialcell[c]{Update\\Normal}}& \textbf{\specialcell[c]{Sub-Contract\\Detection}} & \textbf{\specialcell[c]{Update\\Result}} & \textbf{\specialcell[c]{Normalize\\Result}}\\ \hline
     65,000 & 32 & 11712 & 8296 & 16\\ \hline
     1,01,000 & 32 & 11782 & 8138 & 16\\ \hline
     2,19,500 & 35 & 11713 & 7940 & 17\\ \hline
     3,00,000 & 70 & 11736 & 8094 & 17\\ \hline
     4,66,737 & 73 & 11747 & 7754 & 20\\ \hline
  \end{tabular}
  }
{\caption{\small \specialcell[c]{Execution Time for\\Sub-Contract Algorithm in MySQL}}}
 \ttabbox
  {\tabcolsep=0.06cm 
  \begin{tabular}{|c|c|c|c|c|}
  \hline
      \textbf{\specialcell[c]{Dataset\\Size}} & \multicolumn{4}{|c|}{\textbf{Execution Time(msec)}}\\
    \hline
    & \textbf{\specialcell[c]{Update\\Normal}} & \textbf{\specialcell[c]{Sub-Contract\\Detection}} & \textbf{\specialcell[c]{Update\\Result}} & \textbf{\specialcell[c]{Normalize\\Result}}\\ \hline
     65,000 & 118 & 1542 & 2077 & 5\\ \hline
     1,01,000 & 140 & 1707 & 2773 & 5\\ \hline
     2,19,500 & 202 & 2534 & 2369 & 6\\ \hline
     3,00,000 & 336 & 3442 & 5261 & 9\\ \hline
     4,66,737 & 560 & 4149 & 5334 & 9\\ \hline
  \end{tabular}
  }
{ \caption{\small \specialcell[c]{Execution Time for\\Sub-Contract Algorithm in Neo4j}}}
\end{floatrow}
\end{table}


\begin{figure}
  \centering
  \mbox{
    \subfigure[Execution Time for Sub-Contract Algorithm on MySQL]{\includegraphics[width=7.5cm,height=5cm]{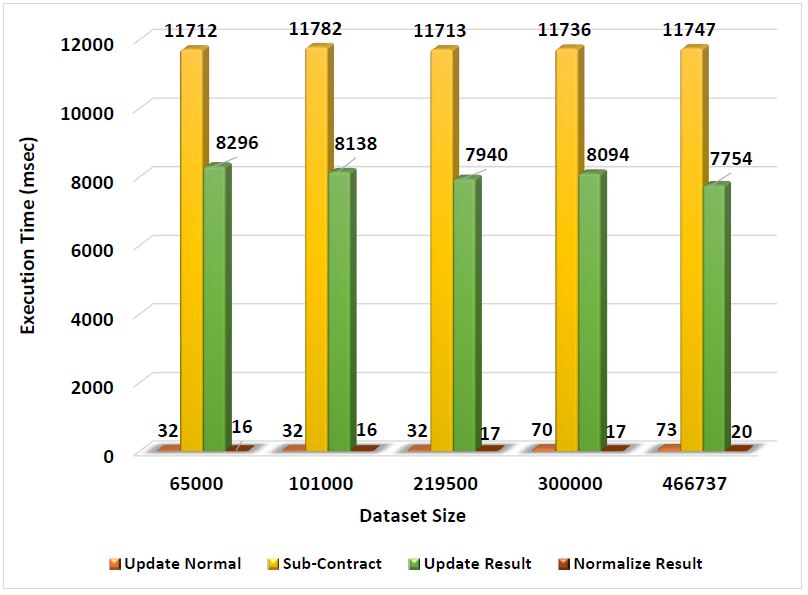}}\quad
    \subfigure[Execution Time for Sub-Contract Algorithm on Neo4j]{\includegraphics[width=7.5cm,height=5cm]{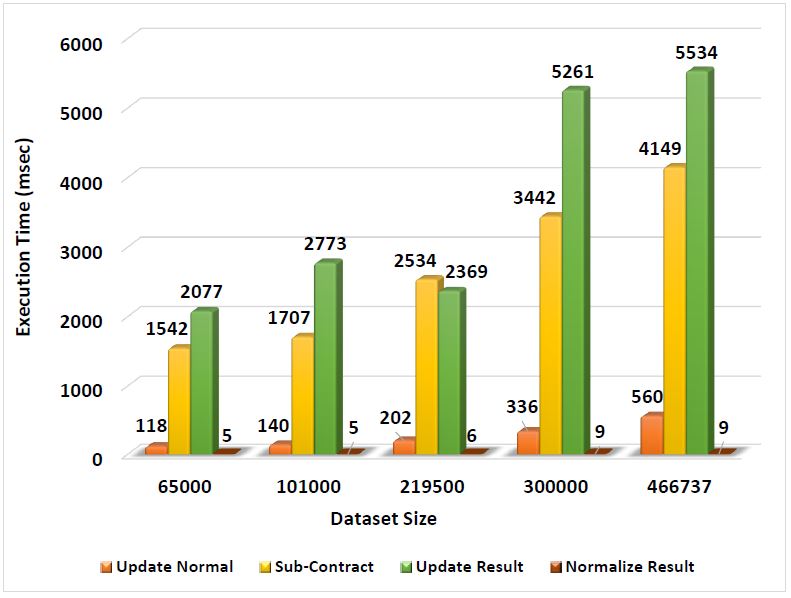}}}
  \caption{Execution Time for Sub-Contract Algorithm}
\end{figure}
\par{ We believe that detecting sub-contracting actors in MySQL is compute intensive and hence time consuming task. The operation is expensive because detecting sub-contracting actors involves retrieving all records for a particular CaseID and then applying self-join on the result set. Joins are compute intensive task in MySQL and involves Cartesian product of the tables based on the condition, followed by selection. On the other hand, detecting sub-contracting actors in Neo4j is equivalent to traversing relationships in Neo4j using index-free adjacency. In our opinion, index-free adjacency achieves better traversal because relationships are stored as first-class citizens in Neo4j and no computation(s) are performed for deriving these relationships. Another major aspect that Fig. 7(a) and Fig. 7(b) brings forward is that write operation in MySQL roughly takes the same amount of time for all dataset sizes. We believe that MySQL needs to write values, albeit zero or default, for all those relations that does not even exist. On the other hand, Neo4j's approach to creating relationship between nodes only when needed is an effcient approach and thus takes less time as compared to MySQL. Although we observe a gradual increasing trend in \textit{Update Result} (or write operation) in Neo4j, we conclude that write operation in Neo4j is linearly scalable with dataset size, On the other hand, write operations in MySQL is fairly constant for all dataset sizes and comparatively higher than Neo4j.}
\renewcommand{\arraystretch}{.5}
\begin{table}
\centering
\begin{tabular}{|c|c|c|c|c|c|}
  \hline
      \textbf{Tables} & \multicolumn{5}{|c|}{\textbf{Dataset Size}}\\
    \hline
    & \textbf{65000}& \textbf{101000} & \textbf{219500}& \textbf{300000} & \textbf{466737}\\ \hline
     Dataset & 4734976 & 6832128 & 13123584 & 18366464 & 27836416\\ \hline
    Organised Data & 4734976 & 6832128 & 13123584 & 18366464 & 27836416\\ \hline
    Result Matrix & 1589248 & 1589248 & 1589248 & 1589248 & 1589248\\ \hline
  \end{tabular}
\caption{\small Disk Space Usage (bytes) for MySQL tables ($Sub-Contract$)}
\end{table}
\par{Table 13 and Fig. 8(a) presents the disk space (in bytes) taken by tables in MySQL. These statistics include both initial tables, intermediate tables, final tables and index, if any. Table 14 and Fig. 8(b) shows the disk usage of various graph elements for the implementation of sub-contraction algorithm for different dataset sizes. The disk space for nodes is contributed by three different nodes type \emph{viz.} Case nodes, Actor nodes and Unique Actor nodes. There are three relationships that contribute to relationship disk space \emph{viz.} $[:CONTAINS]$ relationships that connects Case node to Actor nodes, [:RELATED\textunderscore TO] connects Actor to Actor who satisfy the sub-contraction criteria and $[:SUBCONTRACT]$ connects UNIQUEACTOR nodes to UNIQUEACTOR nodes with the actual sub-contraction value between the unique actors. Readers are suggested to refer to Section 5 for better understanding of the tables and graph elements associated with the implementation of Sub-Contract algorithm in MySQL and Neo4j. We observe that MySQL disk space usage is 30 times lower than Neo4j. }

\begin{table}[H]
\centering
\begin{tabular}{|c|c|c|c|c|c|}
  \hline
      \textbf{Tables} & \multicolumn{5}{|c|}{\textbf{Dataset Size}}\\
    \hline
    & \textbf{65000}& \textbf{101000} & \textbf{219500}& \textbf{300000} & \textbf{466737}\\ \hline
    Nodes & 982212 & 1523732 & 3360798 & 4598454 & 7190330\\ \hline
    Relationships & 153477291 & 183955761 & 285778449 & 375437997 & 490033038\\ \hline
    Properties & 384189475 & 461537287 & 719874720 & 946265404 & 1238579332\\ \hline
  \end{tabular}
\caption{\small Disk Space Usage (bytes) for Neo4j elements ($Sub-Contract$)}
\end{table}
\par{We believe that Neo4j's disk space usage for Neo4j is higher than MySQL because of the number of properties being used to store information used in sub-contract detection. Each property in Neo4j takes 41 bytes and relationship takes 33 bytes. Apart from this, Neo4j stores relationships using index-free adjacency which means every relationship is explicitly stored without any pointers or indexes. This contributes to higher disk usage in Neo4j. On the other hand, MySQL stores information in tables whose size is determined by the data type involved and thus consumes lesser disk space.}

\begin{figure}
  \centering
  \mbox{
    \subfigure[Disk space usage for MySQL tables]{\includegraphics[width=7.5cm,height=5cm]{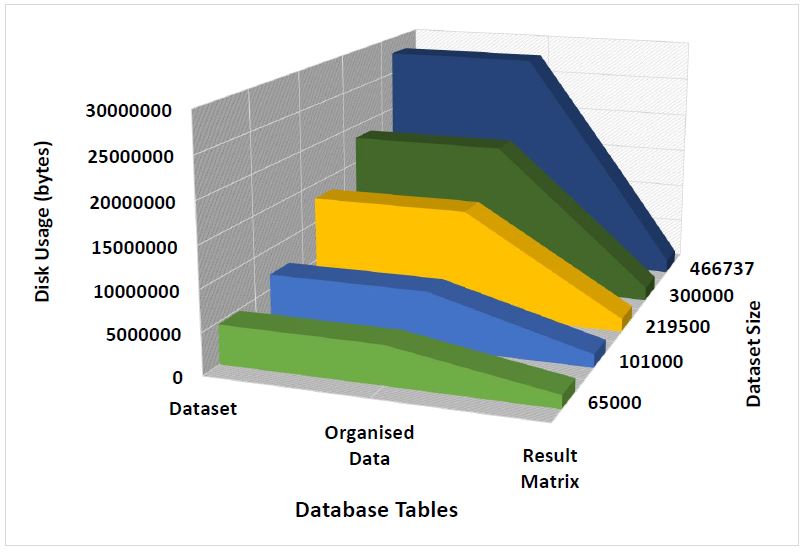}}\quad
    \subfigure[Disk space usagefor Neo4j elements]{\includegraphics[width=7.5cm,height=5cm]{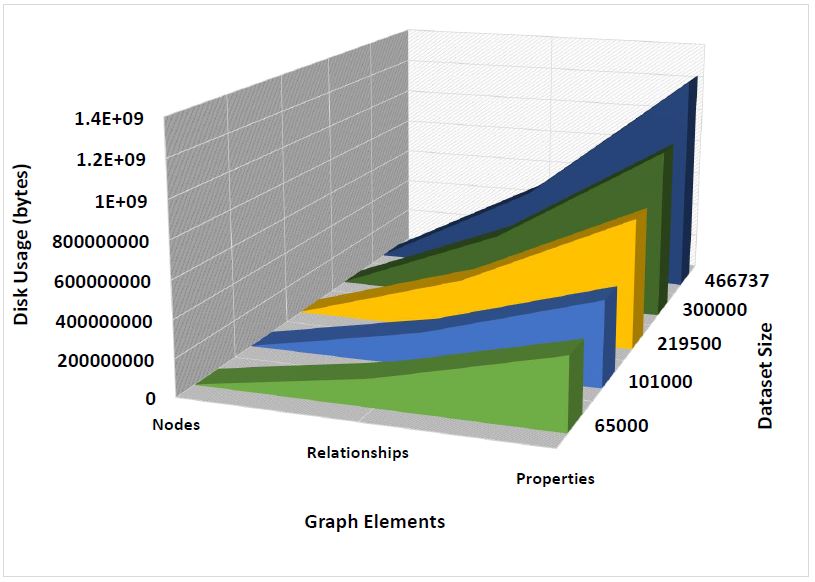}}}
  \caption{Disk Space Usage for Sub-Contract Algorithm}
\end{figure}

\par{We conduct a general experiment to study the variance of memory, disk and process parameters in MySQL and Neo4j using Performance Monitor. We use SQL and CYPHER implementations of Similar-Task algorithm for the experiment. Fig. 9(a) and 9(b) presents bar graphs for various memory, process and disk related parameters. We observe that Neo4j achieves higher level of caching and outperforms MySQL by a factor of 18. Though, Neo4j is seen to incur 3 times more page faults per second, such page faults may not necessarily go to disk. It is further made evident from the fact that MySQL incurs about 6 times more IO operations per second as compared to Neo4j. Fig. 9(b) further strengthens the point with the fact that MySQL spends almost 10 times more time in disk operations and about 6 times more doing disk transfers per second. Based on experimental results, we conclude that Neo4j is more IO efficient than MySQL and with higher physical memory, Neo4j's performance would significantly improve.}
\setlength\intextsep{1.25\baselineskip plus 3pt minus 2 pt}
\begin{figure}[H]
  \centering
  \mbox{
    \subfigure[Statistics for Memory and Process parameters]{\includegraphics[width=7.5cm,height=4.5cm]{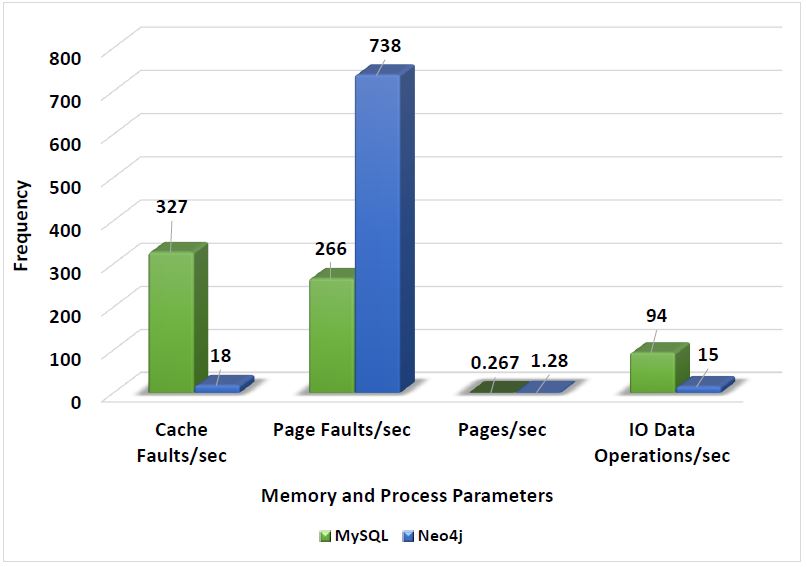}}\quad
    \subfigure[Statistics for Disk parameters]{\includegraphics[width=7.5cm,height=4.5cm]{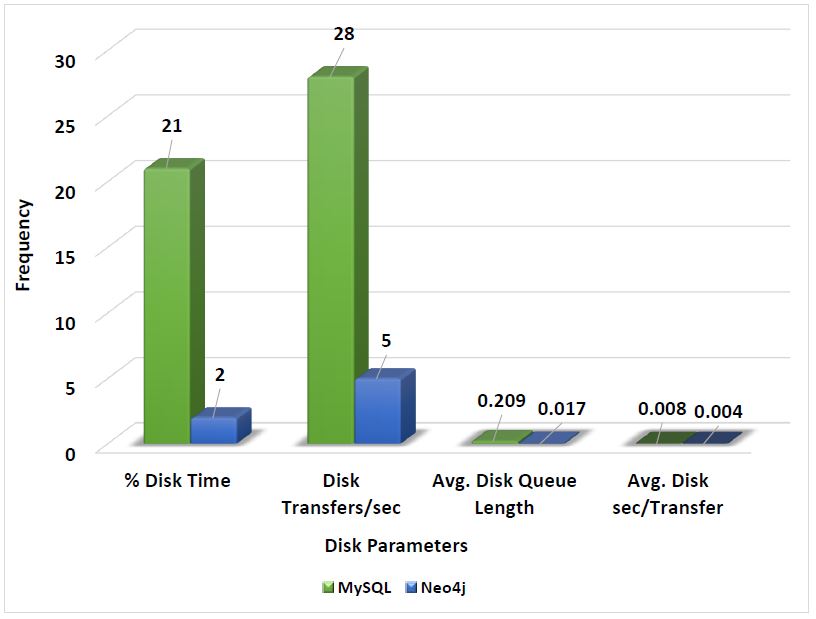}}}
  \caption{Comparison of memory and disk performance monitors for Similar-Task algorithm}
\end{figure}
%

\section{Conclusion}
\par{In this paper, we present the implementation of two different Organizational Mining algorithms in Structured Query Language and CYPHER Query Language. We implement Similar-Task and Sub-Contract algorithm for both native SQL client and as well as by using Java API's using memoization. Furthermore, we benchmark and present performance comparisons of Similar-Task and Sub-Contract algorithms in MySQL and Neo4j. Similar-Task implementation in MySQL is a one-tier application which uses only standard SQL queries and advanced stored procedures. Similarly, implementation in Neo4j is done using standard CYPHER queries. We conclude that Neo4j on an average is 1.25 times faster than MySQL in loading large datasets with only unique elements being created. Based on experimental results, we conclude that similarity calculation in MySQL is 32 times better in MySQL as compared to Neo4j. MySQL outperforms Neo4j in terms of time taken for write operations. The time taken by MySQL is 1.5 times lower as compared to Neo4j. The disk space occupied by elements of graph database in Neo4j is 12 times lower than disk space taken by tables in MySQL. We conclude that Neo4j is more efficient than MySQL in terms of storing only unique information in the database.}
\par{Sub-Contract implementation in MySQL is a one-tier application which also uses standard SQL queries and advanced stored procedures. Similarly, implementation in Neo4j is done using native CYPHER queries. Also, we implement Sub-Contract algorithm with Java API using memoization. We conclude that Neo4j on an average is 1.15 times faster than MySQL in loading large datasets with duplication of elements being allowed. Based on experimental results, we conclude that traversing relationships to find sub-contracting actors in Neo4j is 7 times better as compared to MySQL. Neo4j outperforms MySQL in terms of time taken for write operations. The time taken by Neo4j is 4 times lower as compared to MySQL. However, disk space taken by graph elements in Neo4j is over 30 times higher as compared to MySQL because of the need to store redundant information.}

\par{In general, we conclude that Neo4j gives better performance than MySQL in loading large datasets with performance benefits of upto 25 percent. Tasks which involve traversing relationships followed by computation (like Similar-Task algorithm) are time consuming in Neo4j. However, Neo4j performs better than MySQL when finding relationship is concerned (like Sub-Contract algorithm) and is seen to perform 7 times better than MySQL. Also Neo4j gives better write time performance as volume of data increases. Based on our analysis of resources during experiments of Similar-Task algorithm, we conclude that Neo4j achieves higher level of caching and incurs almost 6 times lower disk IO operations. Our analysis reveals that Neo4j spends 10x less time doing disk operations with an average of 6 times lower disk transfers per second.}
\nocite{*}
\bibliographystyle{unsrt}
\bibliography{citations}
\end{document}